\documentclass[twocolumn,preprintnumbers,amsmath,superscriptaddress,amssymb,aps]{revtex4-1}

\usepackage{graphicx}
\usepackage{dcolumn}
\usepackage{bm}
\usepackage{amsmath}
\usepackage{amssymb}
\usepackage{graphicx}
\usepackage{indentfirst}
\usepackage{booktabs}
\usepackage{multirow}
\usepackage{colortbl}

\selectfont

\begin{document}
\title{Multifield tunable valley splitting and anomalous valley Hall effect in two-dimensional antiferromagnetic MnBr}

\author{Yiding Wang}
\thanks{These authors contributed equally to this work.}
\address{School of Physics and Mechatronics Engineering, Jishou University, Jishou, Hunan, 416000, People's Republic of China}
\address{Department of Physics and Electronic Engineering, Tongren University, Tongren 554300, People's Republic of China}
\address{State Key Laboratory for Mechanical Behavior of Materials, School of Materials Science and Engineering, Xi'an Jiaotong University, Xi'an, Shaanxi, 710049, People's Republic of China}
\author{Hanbo Sun}
\thanks{These authors contributed equally to this work.}
\address{State Key Laboratory for Mechanical Behavior of Materials, School of Materials Science and Engineering, Xi'an Jiaotong University, Xi'an, Shaanxi, 710049, People's Republic of China}
\author{Chao Wu}
\address{State Key Laboratory for Mechanical Behavior of Materials, School of Materials Science and Engineering, Xi'an Jiaotong University, Xi'an, Shaanxi, 710049, People's Republic of China}
\author{Weixi Zhang}
\email{zhangwwxx@sina.com}
\address{Department of Physics and Electronic Engineering, Tongren University, Tongren 554300, People's Republic of China}
\author{San-Dong Guo}
\address{School of Electronic Engineering, Xi'an University of Posts and Telecommunications, Xi'an, 71021, People's Republic of China}
\author{Yanchao She}
\email{ycshe@xtu.edu.cn}
\address{School of Physics and Mechatronics Engineering, Jishou University, Jishou, Hunan, 416000, People's Republic of China}
\address{Department of Physics and Electronic Engineering, Tongren University, Tongren 554300, People's Republic of China}
\author{Ping Li}
\email{pli@xjtu.edu.cn}
\address{State Key Laboratory for Mechanical Behavior of Materials, School of Materials Science and Engineering, Xi'an Jiaotong University, Xi'an, Shaanxi, 710049, People's Republic of China}
\address{State Key Laboratory of Silicon and Advanced Semiconductor Materials, Zhejiang University, Hangzhou, 310027, People's Republic of China}
\address{State Key Laboratory for Surface Physics and Department of Physics, Fudan University, Shanghai, 200433, People's Republic of China}

\date{\today}

\begin{abstract}
Compared to the ferromagnetic materials that realize the anomalous valley Hall effect by breaking time-reversal symmetry and spin-orbit coupling, the antiferromagnetic materials with the joint spatial inversion and time-reversal ($\emph{PT}$) symmetry are rarely reported that achieve the anomalous valley Hall effect. Here, we predict that the antiferromagnetic monolayer MnBr possesses spontaneous valley polarization. The valley splitting of valence band maximum is 21.55 meV at K and K' points, which is originated from Mn-d$_{x^2-y^2}$ orbital by analyzing the effective Hamiltonian. Importantly, monolayer MnBr has zero Berry curvature in the entire momentum space but non-zero spin-layer locked Berry curvature, which offers the condition for the anomalous valley Hall effect. In addition, the magnitude of valley splitting can be signally tuned by the onsite correlation, strain, magnetization rotation, electric field, and built-in electric field. The electric field and built-in electric field induce spin splitting due to breaking the $\emph{P}$ symmetry. Therefore, the spin-layer locked anomalous valley Hall effect can be observed in MnBr. More remarkably, the ferroelectric substrate Sc$_2$CO$_2$ can tune monolayer MnBr to realize the transition from metal to valley polarization semiconductor. Our findings not only extend the implementation of the anomalous valley Hall effect, but also provides a platform for designing low-power and non-volatile valleytronics devices.
\end{abstract}

\maketitle
\section{Introduction}
Valley degrees of freedom is the third degree of freedom for electrons outside of charge and spin, which has recently attracted extensive attention since it provides remarkable opportunities for realizing the next-generation of ultra-fast speed, ultra-high capacity, low power consumption, and non-volatile devices \cite{1,2,3,4,5,6,7,8}. The valley refers to a local energy maximum or minimum point in the valence or conduction band, which the energy extremum is robust against impurity scattering and phonon due to the large separation in the momentum space \cite{9}. The current focus of valley investigation is understanding how to stably manipulate the valley degrees of freedom, thereby generating robust valley polarization \cite{10,11,12,13,14,15,16,17,18,19}. Physically, there are two strategies to realize valley polarization. One is optical excitation with circularly polarized light \cite{20,21}, the other is the symmetry breaking \cite{22,23}. The former is a dynamic process, while the latter is especially concerned, particularly the intrinsic valley polarization.

There are also two kinds of ways to achieve the intrinsic valley polarization. One is to break the symmetry of spatial inversion ($\emph{P}$) via ferroelectric polarization \cite{16,24,25}, and the other broke the time-reversal symmetry ($\emph{T}$) through magnetism \cite{10,11,12,13,14,15,16,17,18,19}. This kind of material is named ferrovalley material \cite{10}. Until now, the ferrovalley via breaking $\emph{T}$ symmetry is mainly focused on ferromagnetic (FM) materials. Compared with the FM materials, the antiferromagnetic (AFM) material is robust against external magnetic perturbation, has high storage density, and possesses ultrafast writing speed (about three orders of magnitude higher than the FM material) due to the zero magnetic moment \cite{26,27}. It is well-known that the two sublattices have opposite spin vectors for the AFM hexagonal lattice. Both the $\emph{P}$ and $\emph{T}$ symmetries are broken, but the joint symmetry $\emph{PT}$ is retained. Therefore, the AFM ferrovalley provides inviting potential for the spintronic and valleytronic applications. Unfortunately, the spontaneous valley polarization in AFM materials is exceedingly rare, which makes the circumstances worse. Simultaneously, the anomalous valley Hall effect is undesirably suppressed for the AFM materials.

In this work, based on the density functional theory (DFT) and Hamiltonian model, we propose that the AFM monolayer MnBr is a highly fascinating candidate for the abundant valley contrasting physics. Our results show that monolayer MnBr is a semiconductor with the spin up and spin down bands degeneracy without considering spin-orbit coupling (SOC). When the SOC is included, the valence band maximum (VBM) has 21.55 meV valley splitting at K and K' points, which originated from Mn-d$_{x^2-y^2}$ orbital by analyzing the effective Hamiltonian. Importantly, the value of valley splitting can be tuned not only by the onsite correlation, strain, magnetization rotation, electric field, and built-in electric field. But also the phase transition of valley polarized semiconductor to metal can be controlled by ferroelectric substrate Sc$_2$CO$_2$. Our findings open an avenue for the investigation of valley physical quantity in the AFM material, realizing the energy conservation, fast operating spintronic and valleytronic devices.

\section{COMPUTATIONAL METHODS}
Based on the framework of the DFT, we investigate the electronic and magnetic properties using the Vienna $Ab$ $initio$ simulation package (VASP) \cite{28,29}. The generalized gradient approximation (GGA) with the Perdew-Burke-Ernzerhof (PBE) functional is employed to describe the exchange-correlation energy \cite{30}. The plane-wave basis with a kinetic energy cutoff of 500 eV is used. A vacuum of 30 $\rm \AA$ is added along the $\emph{c}$-axis, to avoid the interaction between the sheet and its periodic images. The convergence criteria of the force and the total energy are set to -0.005 eV/$\rm \AA$ and 10$^{-6}$ eV, respectively. To describe strongly correlated 3$\emph{d}$ electrons of Mn, the GGA+U method is adopted with the effective U value (U$_{eff}$ = U - J) of 4 eV \cite{31,32,33}. The zero damping DFT-D3 method of Grimme is considered for the Van der Waals (vdW) correction in MnBr/Sc$_2$CO$_2$ heterostructure \cite{34}. To explore the dynamical stability, the phonon spectra are calculated by the PHONOPY code using a finite displacement approach with a 4$\times$4$\times$1 supercell \cite{35}. Thermal stability is evaluated using $\emph{Ab initio}$ molecular dynamic (AIMD) with a 3$\times$3$\times$1 supercell \cite{36}. Moreover, to investigate the Berry curvature, the maximally localized Wannier functions (MLWFs) are performed to construct an effective tight-binding Hamiltonian by Wannier90 code \cite{37,38}.

\section{RESULTS AND DISCUSSION }	
\subsection{Structure and stability}
As shown in Fig. 1(a, d), it exhibits the crystal structure of monolayer MnBr. The monolayer MnBr possesses a two-dimensional (2D) hexagonal lattice and consists of septuple layers of Br-Mn-Mn-Br. There are three Br and three Mn atoms nearest to each Mn atom, forming an octahedral crystal field. The space group of monolayer MnBr is P$\bar{3}$m1. The lattice constant of MnBr is optimized to 3.60 $\rm \AA$. To evaluate the dynamic stability, the phonon spectra are investigated. As shown in Fig. 1(c), the absence of imaginary frequencies verifies that monolayer MnBr is dynamically stable. To explore the bonding features of monolayer MnBr, we calculated its electron localization function (ELF). As shown in Fig. 1(e), the electrons are mainly localized between Mn and Mn atoms and around the Br atoms, showing typical covalent bonding for the Mn-Mn bond and ionic bonding for the Mn-Br bond. Moreover, the thermodynamic stability is also estimated by the AIMD calculation. As shown in Fig. 1(f), the total energies of monolayer MnBr fluctuate very small evolving during 5 ps at 300 K, indicating that monolayer MnBr has good thermal stability.

\begin{figure}[htb]
\begin{center}
\includegraphics[angle=0,width=1.0\linewidth]{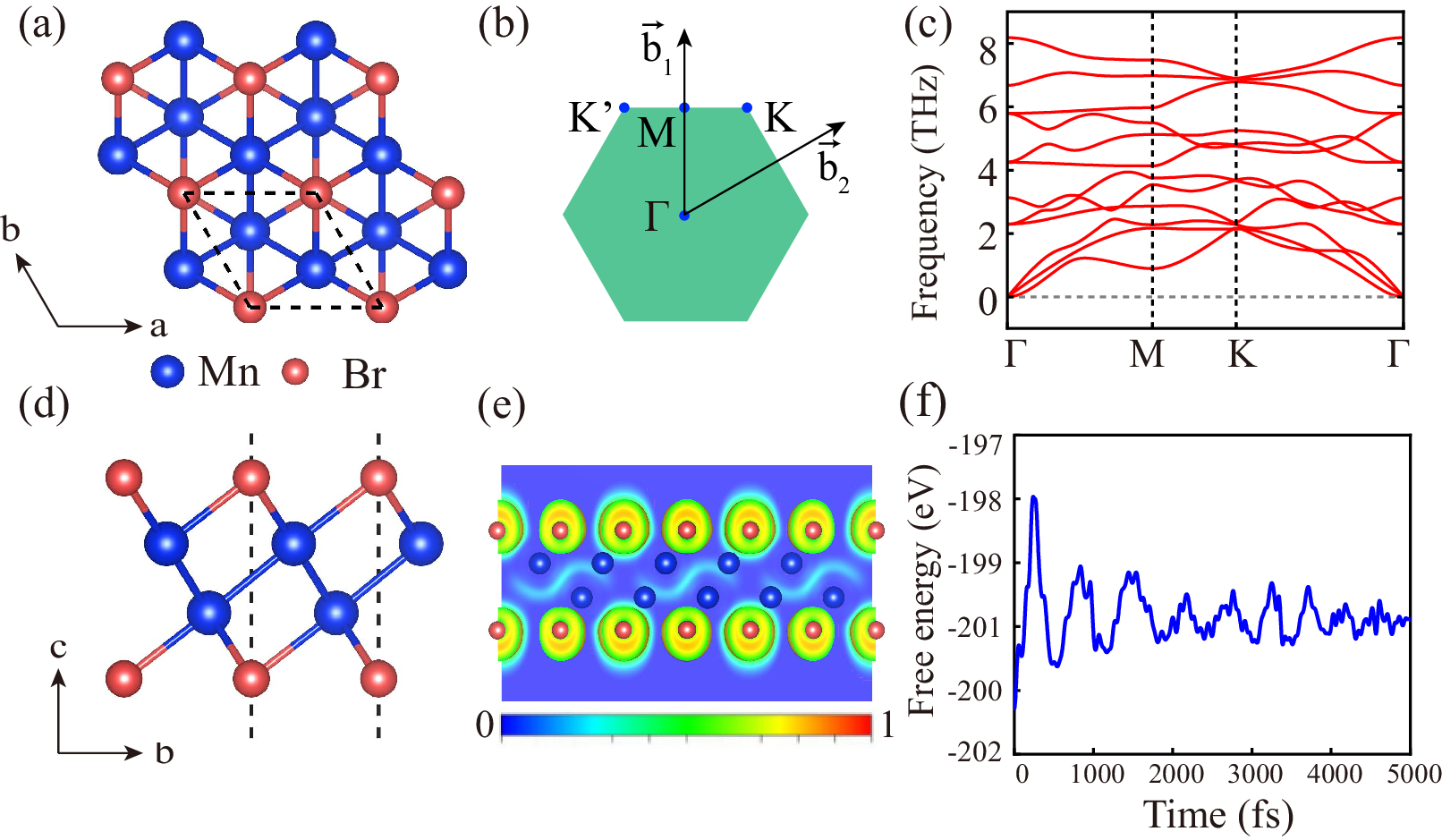}
\caption{(a, d) The top and side views of the crystal structure for monolayer MnBr, respectively. The blue and pink balls represent Mn, and Br elements, respectively. (b) The Brillouin zone (BZ) of the honeycomb lattice with the reciprocal lattice vectors $\vec{b}_1$ and $\vec{b}_2$. The $\Gamma$, K, K', and M are high-symmetry points in the BZ. (c) The calculated phonon dispersion curves along the high-symmetry lines with the first BZ of MnBr. (e) Electron localization function of monolayer MnBr. (f) Total energy fluctuations of monolayer MnBr during 5 ps AIMD simulation at 300 K.
}
\end{center}
\end{figure}

\subsection{Magnetic property}
The valence electronic configuration of the isolated Mn atom is 3d$^5$4s$^2$. For the MnBr crystal, the Mn atom would lose three electrons to the three neighboring Br atoms due to its surroundings, resulting in the electronic configuration of 3d$^4$4s$^0$. In the octahedral crystal field, the $\emph{d}$ orbitals is split into two groups: t$_{2g}$ (d$_{xy}$, d$_{yz}$, d$_{xz}$ orbital), and e$_g$ (d$_{x^2-y^2}$, d$_{z^2}$ orbital). According to the Pauli exclusion principle and Hund's rule, the electronic configuration of Mn would half fill the t$_{2g}$ orbitals and half fill one of two e$_g$ orbitals. Therefore, the magnetic moment of each Mn atom is 4 $\mu_B$.

\begin{figure}[htb]
\begin{center}
\includegraphics[angle=0,width=1.0\linewidth]{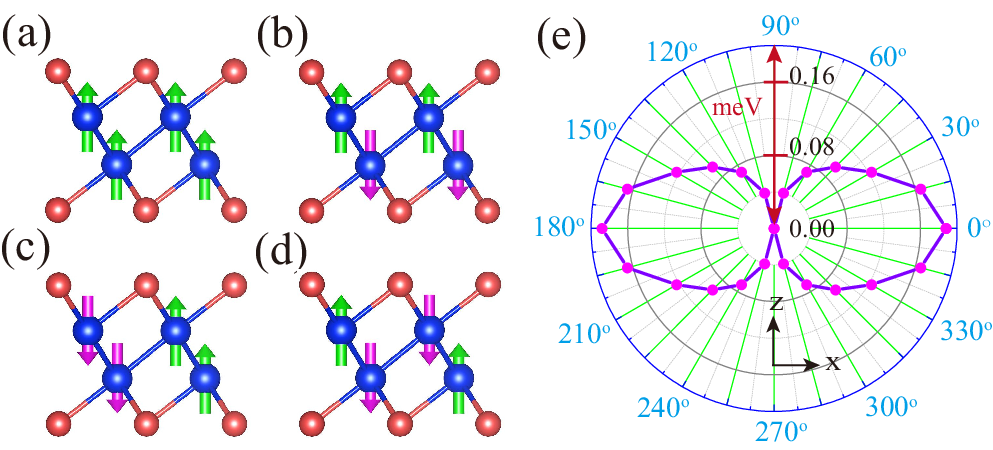}
\caption{(a-d) are FM, AFM1, AFM2, and AFM3 configurations for monolayer MnBr, respectively. (e) Angular dependence of the MAE of monolayer MnBr with the direction of magnetization lying on the xz plane.
}
\end{center}
\end{figure}

To confirm the magnetic ground state of monolayer MnBr, as shown in Fig. 2(a-d), four possible magnetic configurations including the FM, AFM1, AFM2 and AFM3 are considered. We calculate the total energy difference between FM, AFM1, AFM2, and AFM3 using the GGA + U method. According to previous reports \cite{30,31,32}, the selection of U value for Mn atom in 2D materials, we choose U$_{\rm eff}$ = 4 eV to investigate the monolayer MnBr. The AFM1 state is 4.414 eV, 3.305 eV, and 2.235 eV lower in energy than the FM, AFM2, and AFM3 states, respectively.

For 2D magnetic materials, the out-of-plane is the direction of easy magnetization that is the basis for all research. Therefore, we investigate the magnetic anisotropy energy (MAE), which is defined as MAE = E$_{100}$ - E$_{001}$, where E$_{100}$ and E$_{001}$ denote the total energy of the magnetic moment along [100] and [001] direction, respectively. The MAE is 0.18 meV, showing the magnetization along the [001] direction. In addition, for the octahedral crystal field of monolayer MnBr, the angular dependence of the MAE can be written employing the equation
\begin{equation}
\rm MAE= K_1 cos^2\theta + K_2 cos^4\theta,
\end{equation}
where $\theta$ is the azimuthal angle of rotation and K$_1$ and K$_2$ are the anisotropy constants.
If K$_1$ $<$ 0, the benefited magnetization direction will be along the out-of-plane (z-axis), while K$_1$ $>$ 0 indicates that it will be parallel to in-plane (x-axis). The MAE of monolayer MnBr exhibits a good fit of Eq (3) as appeared in Fig. 2(e), it indicates that the MAE strongly depends on the direction of magnetization in the xz plane.

\subsection{Band structure and anomalous valley Hall effect}
Then, we concentrate on band structure and associated valley splitting of monolayer MnBr. When the SOC is switched off, as shown in Fig. 3(a), the spin up and spin down bands are degenerate. Moreover, it is an indirect band gap semiconductor with a 1.89 eV band gap. The VBM shows at the K/K' point and they're degenerate, while the conduction band minimum (CBM) exhibits M point. Fig. 3(b, c) show the orbital-resolved band structure, we find that the VBM bands are dominated by Mn d$_{x^2-y^2}$+d$_{z^2}$ orbitals, while the CBM bands are mainly contributed by Mn d$_{xy}$+d$_{yz}$+d$_{xz}$ orbitals. In the existence of SOC, as shown in Fig. 3(d), the valley degeneracy of K and K' at the VBM bands disappears. Simultaneously, the valley polarization is spontaneously induced, and the valley splitting is 21.55 meV. For the orbital-resolved band structure, as shown in Fig. 3(e, f), they are completely consistent with the considering SOC and without the considering SOC.

\begin{figure}[htb]
\begin{center}
\includegraphics[angle=0,width=1.0\linewidth]{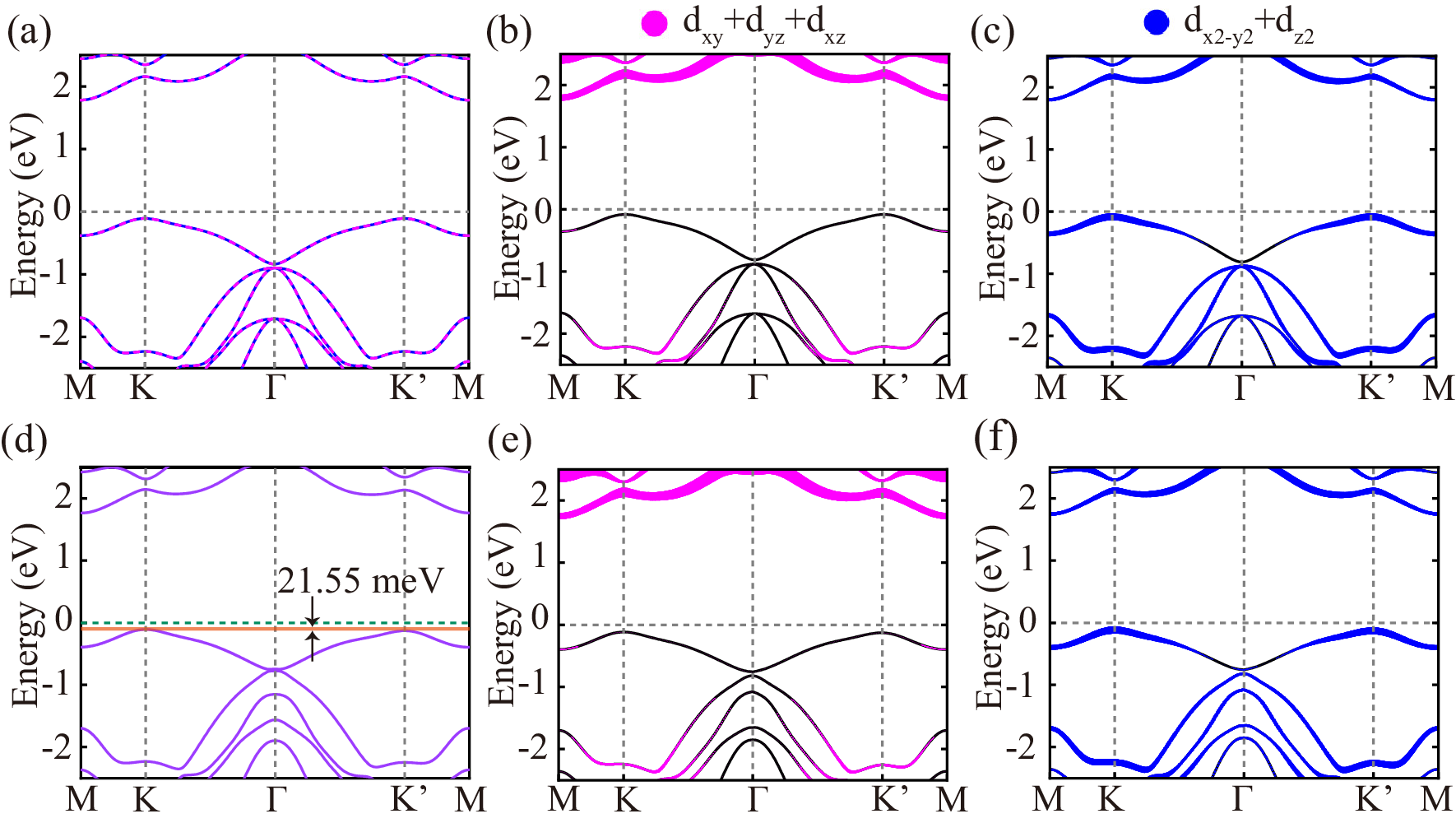}
\caption{Band structures of monolayer MnBr (a) without and (d) with considering the SOC. Band structures of monolayer MnBr without considering SOC projected on the (b) d$_{xy}$+d$_{yz}$+d$_{xz}$ and (c) d$_{x^2-y^2}$+d$_{z^2}$ orbitals and with considering SOC projected on the (e) d$_{xy}$+d$_{yz}$+d$_{xz}$ and (f) d$_{x^2-y^2}$+d$_{z^2}$ orbitals. The magenta dotted line and blue line show spin down and spin up bands, respectively. The valence band valley splitting is shown by the orange shading.
}
\end{center}
\end{figure}

To understand the underlying mechanism for the ferrovalley effect in monolayer MnBr, we adopted $|$$\psi$$_v$$^{\tau}$$\rangle$=$|$d$_{x2-y2}$$\rangle$)$\otimes$$|$$\uparrow$$\rangle$ to construct an effective Hamiltonian. It is well known that the orbital angular momentum of the d$_{z^2}$ orbital is zero. Thus, we don't consider the d$_{z^2}$ orbital, since the d$_{z^2}$ orbital doesn't cause valley splitting. Based on the SOC effect as a perturbation term, the effective Hamiltonian can be written as
\begin{equation}
\hat{H}_{SOC} = \lambda \hat{S} \cdot \hat{L} = \hat{H}_{SOC}^{0} + \hat{H}_{SOC}^{1},
\end{equation}
where $\hat{S}$ and $\hat{L}$ are spin angular and orbital angular operators, respectively. The $\hat{H}_{SOC}^{0}$ and $\hat{H}_{SOC}^{1}$ stands for the interaction between the same spin states and between opposite spin states, respectively. Here, we only consider spin up band. Therefore, the $\hat{H}_{SOC}^{1}$ term can be ignored. Consequently, the $\hat{H}_{SOC}^{0}$ can be written by polar angles
\begin{equation}
\hat{H}_{SOC}^{0} = \lambda \hat{S}_{z'}(\hat{L}_zcos\theta + \frac{1}{2}\hat{L}_+e^{-i\phi}sin\theta + \frac{1}{2}\hat{L}_-e^{+i\phi}sin\theta),
\end{equation}
In the out-of-plane magnetization situation, $\theta$ = $\phi$ = 0$^ \circ$, then the $\hat{H}_{SOC}^{0}$ term can be simplified as
\begin{equation}
\hat{H}_{SOC}^{0} = \lambda \hat{S}_{z} \hat{L}_z,
\end{equation}
The energy level of the valley for the VBM can be described as E$_v$$^ \tau$ = $\langle$ $\psi$$_v$$^ \tau$ $|$ $\hat{H}$$_{SOC}^{0}$ $|$ $\psi$$_v$$^ \tau$ $\rangle$. Then, the valley splitting can be described as
\begin{equation}
E_{v}^{K} - E_{v}^{K'} = \langle d_{x^2-y^2} | \hat{H}_{SOC}^{0} | d_{x^2-y^2} \rangle \approx \beta,
\end{equation}
where the $\beta = \lambda \langle d_{x2-y2} |\hat{S}_{z'}| d_{x2-y2} \rangle$. The analytical model confirms that the valley splitting is consistent with our DFT result ($E_{v}^{K}$ - $E_{v}^{K'}$ = 21.55 meV).

\begin{figure}[htb]
\begin{center}
\includegraphics[angle=0,width=1.0\linewidth]{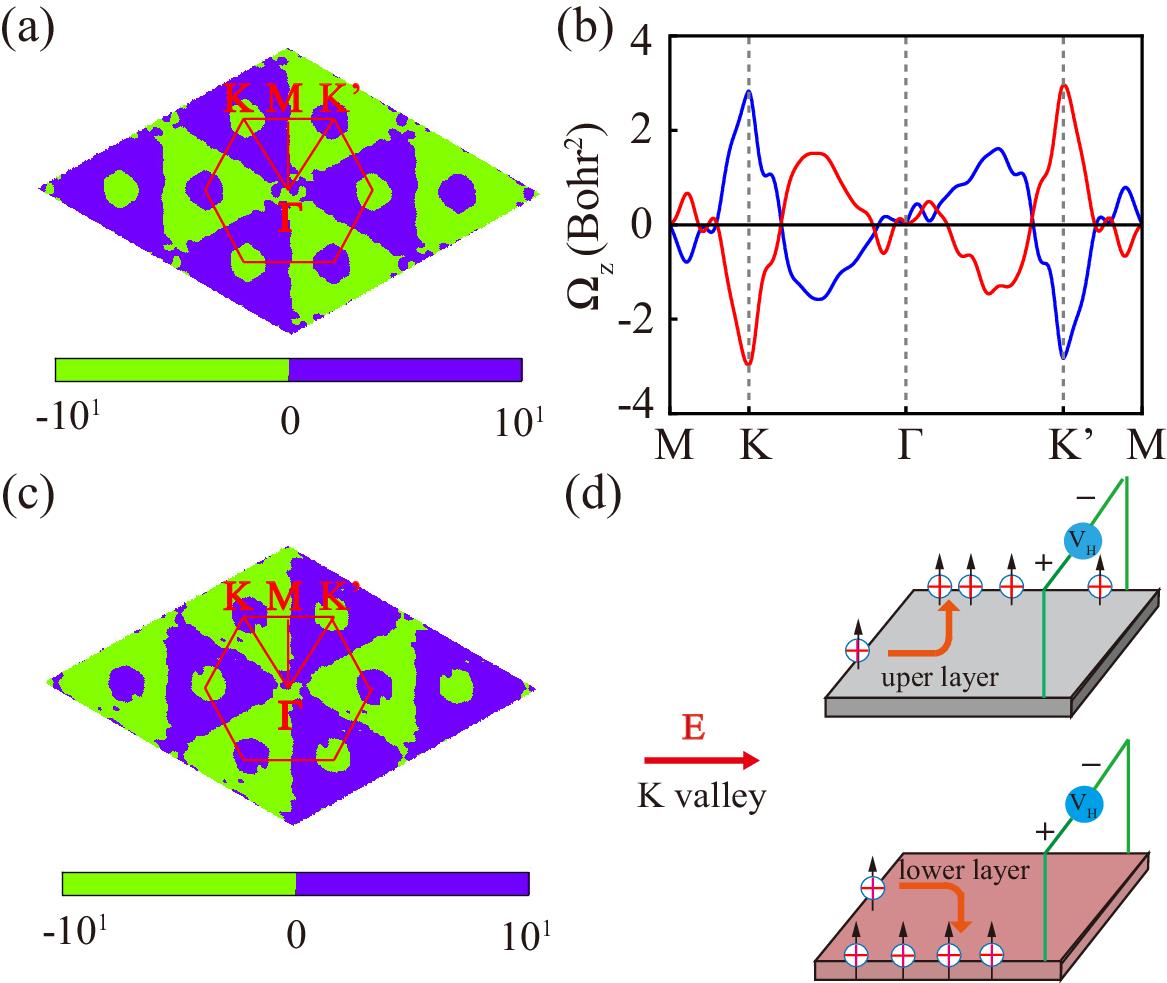}
\caption{The Berry Curvatures of monolayer MnBr for (a) spin up and (c) spin down in the entire BZ region. (b) The Berry Curvature along the high symmetry line. The blue and red lines show spin up and spin down Berry Curvatures, respectively. (d) Schematic diagram of valley layer-spin Hall effect. The holes are represented by the + symbol. Upward arrows and downward arrows on behalf of the spin up and spin down carriers, respectively.
}
\end{center}
\end{figure}

To describe the valley contrasting physics in monolayer MnBr, we calculate its Berry curvature. Due to the joint PT symmetry in monolayer MnBr, the Berry curvature is zero in entire the BZ region. However, monolayer MnBr has two layers of Mn atoms. For each layer of Mn atoms, the PT symmetry is broken. Therefore, the AFM1 state generates spin-layer locking, which makes the Berry curvature of spin up and spin down is opposite sign and equal in magnitude. Fig. 4(a-c) shows the Berry curvature of spin up and spin down in the entire 2D BZ region and along the high-symmetry line. The Berry curvature at the K and K' points have equal magnitude, while it exhibits opposite signs for the same valley of different spin channels and different valleys of the same spin channel. It indicates that produces spin-layer locking Berry curvature and the typical valley feature. While the Fermi level is moved between the K and K' valleys in the VBM band, the spin up and spin down holes of the K valley will be produced and accumulate on the opposite boundary of different layers. As shown in Fig. 4(d), we name this phenomenon the valley layer-spin Hall effect, but the anomalous valley Hall effect is inexistent due to spin degeneracy.

\subsection{Onsite correlation tune valley splitting}
To observe how the band structure and valley splitting, when the strength of onsite correlation Hubbard U is varied. Firstly, we confirm the magnetic ground state of Hubbard U at 0 $\sim$ 5 eV. As shown in Fig. 5(a), the AFM1 state is much lower than the other three magnetic configurations, indicating that the AFM1 is robust. Besides, we explore the MAE at varied Hubbard U. As shown in Fig. 5(b), the variation range is only 0.08 $\sim$ 0.24 meV, showing that the U value has little influence on MAE. Importantly, the directions of easy magnetization are all out-of-plane at varied Hubbard U. On this basis, we calculate the band structures without and with the SOC (see Fig. S1). We can clearly observe that the U value has little effect on the band structure. In the absence of SOC, spin up and spin down bands are still degenerate. Meanwhile, the VBM and CBM remain at K/K' and M points, respectively. When the SOC is included, the valley split line increases from 11.85 to 24.82 meV. In addition, the band gap is increased from 1.40 eV to 1.87 eV and then down to 1.84 eV. In a word, the band gap change is not obvious.

\begin{figure}[htb]
\begin{center}
\includegraphics[angle=0,width=1.0\linewidth]{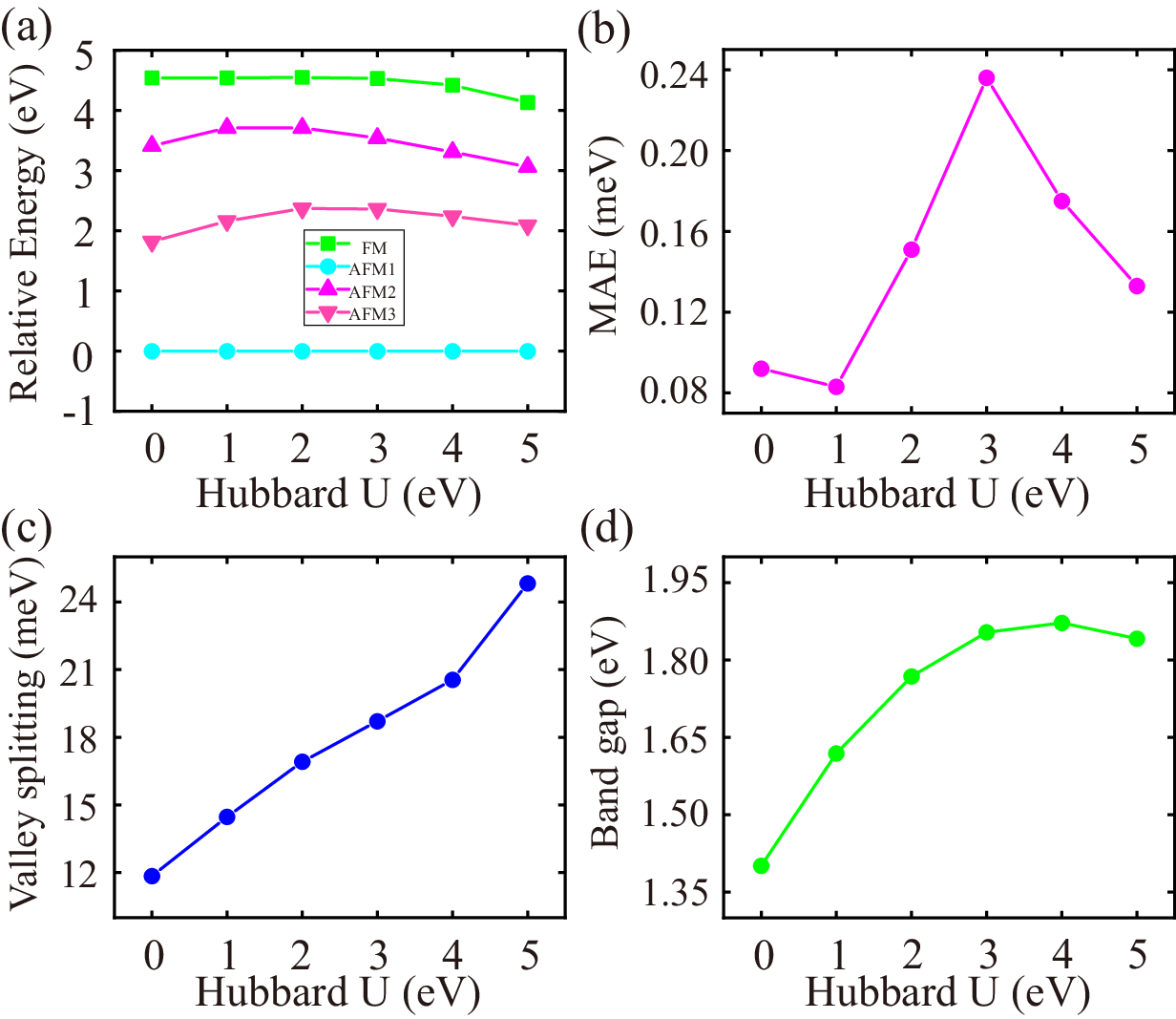}
\caption{ (a) The total energies of monolayer MnBr different magnetic configurations with the different Hubbard U values, which are defined relative to that of the AFM1 state. (b) The MAE with the different Hubbard U values. (c) The valley splitting with the different Hubbard U values. (d) The global band gap with the different Hubbard U values.
}
\end{center}
\end{figure}

\subsection{Strain tune valley splitting}

\begin{figure}[htb]
\begin{center}
\includegraphics[angle=0,width=1.0\linewidth]{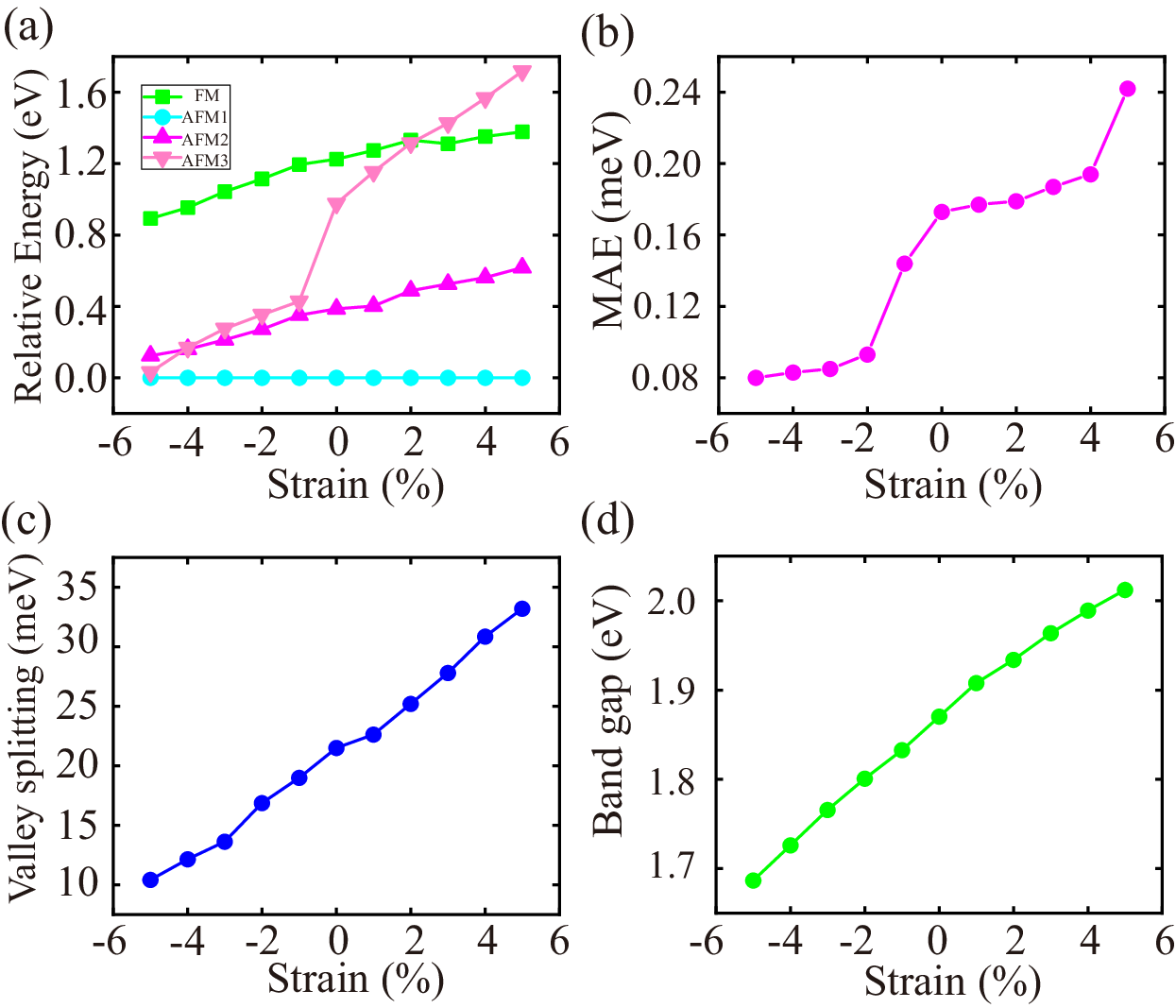}
\caption{ (a) The total energies of monolayer MnBr different magnetic configurations as a function of strain, which are defined relative to that of the AFM1 state. (b) The MAE as a function of strain. (c) The valley splitting as a function of strain. (d) The global band gap as a function of strain.
}
\end{center}
\end{figure}

In device manufacturing, 2D materials are usually supported by a substrate, which may introduce strain to the 2D material due to lattice mismatches \cite{39,40,41}. Therefore, it is very important that the properties of 2D materials have against strain. In the following, we are mainly concerned about the effect of strain on valley splitting. In the calculations, the biaxial strain is defined as $\varepsilon$ = (a-a$_0$)/a$_0$$\times$100$\%$, where a and a$_0$ denote the lattice constant after and before in-plane biaxial strain is applied, respectively. We explore the in-plane biaxial strain in the reasonable range of -5$\%$ $\thicksim$ 5$\%$. As shown in Fig. 6(a), the magnetic ground state is always AFM1 at -5$\%$ $\thicksim$ 5$\%$. It shows that AFM1 is very stable under strain. Besides, as shown in Fig. 6(b), the MAE increased from 0.08 meV to 0.24 meV under strain. And all easy magnetization directions are out-of-plane. The band structures under various strains are shown in Fig. S2 and Fig. S3. As shown in Fig. 6(c), we find that the valley splitting increases from 10.42 meV at -5 $\%$ strain to 33.21 meV at 5 $\%$ strain. It means that the strain is significantly adjustable to the valley splitting. In addition, the band gap has also increased from 1.69 eV to 2.01 eV [see Fig. 6(d)]. Interestingly, the variation trend of valley splitting and band gap is consistent with the MAE. This finding means that the magnetic ground state of monolayer MnBr not only has strain resistance, but also valley splitting can be effectively tuned by strain.

\subsection{Magnetic tune valley splitting}
Considering that monolayer MnBr has both AFM and ferrovalley properties, which is a typical multiferroic material. We are very curious about the coupling strength between multiferroic orders. Hence, we explore the effect of the magnetization direction on the valley splitting. As shown in Fig. 7, it shows the valley splitting of the valence band as a function of the magnetization direction. When the magnetization direction rotates from the in-plane (0$^ \circ$) to the out-of-plane (90$^ \circ$), the valley splitting of valence continuously increases from 0.91 meV to 21.55 meV. If the magnetization direction continues to rotate to the -x axis (180$^ \circ$), the value of valley splitting continuously decreases from the maximum to 0.91 meV. Interestingly, the variation trend of valley splitting is opposite to that of the MAE. Namely, the valley splitting is the smallest when the MAE is maximum, while the valley splitting is the largest as the MAE is minimal. The valley splitting exhibits the twofold relationship, which further confirms that the established effective Hamiltonian of valley splitting is correct.

\begin{figure}[htb]
\begin{center}
\includegraphics[angle=0,width=0.7\linewidth]{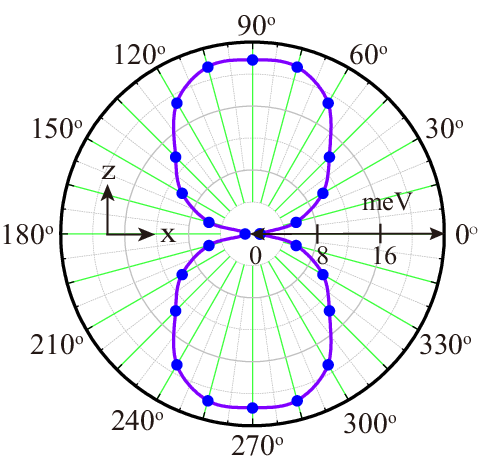}
\caption{ Angular dependence of the valence band valley splitting of monolayer MnBr with the direction of magnetization lying on the xz plane is shown by the blue line.
}
\end{center}
\end{figure}

\subsection{Electric field tune valley splitting}
From the device application point of view, the electric field is the most effective means of tuning the physical quantity \cite{41,42}. Most importantly, the electric fields enable fast, non-damaging, and efficient regulation of the device. Therefore, we investigate the valley splitting with the electric field 0.0$\thicksim$ 0.3 V/$\rm \AA$. An out-of-plane electric field breaks the $\emph{P}$ symmetry. As a result, the $\emph{PT}$ symmetry of monolayer MnBr system disappears. Eventually, the spin splitting is caused. Importantly, the magnetic ground state is still AFM1 configuration, and the direction of easy magnetization is also along out-of-plane. As shown in Fig. 8(a), we can clearly observe that the electric field can effectively tune the value of valley splitting. The valley splitting reduced from 21.55 meV at 0.0 V/$\rm \AA$ to 20.70 meV at 0.3 V/$\rm \AA$. Simultaneously, the global band gap reduced from 1.870 eV to 1.854 eV. Fig. 8(c) shows the band structure of the Mn layer resolved with SOC at E = 0.3 V/$\rm \AA$. It can be clearly observed Mn layer splitting of the band structure under the external electric field. Hence, in the hole-doping case, the spin down holes of K valley will be produced and accumulate on one boundary of the lower layer under an in-plane electric field [see Fig. 4(d)]. Consequently, the spin-layer locked anomalous valley Hall effect will be observed. The characteristic exhibits that monolayer MnBr is an ideal candidate for the high performance valleytronic devices.

\begin{figure}[htb]
\begin{center}
\includegraphics[angle=0,width=1.0\linewidth]{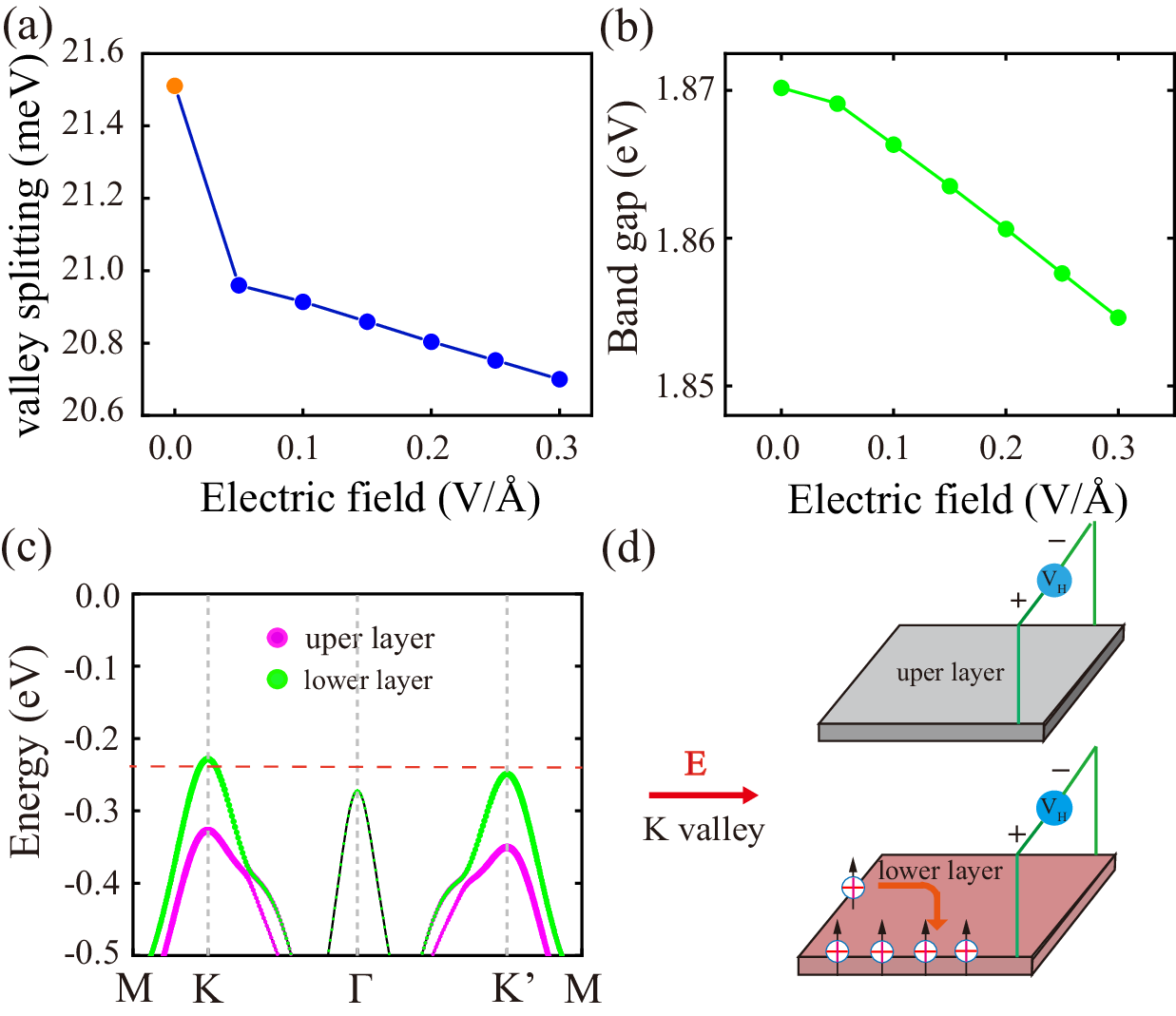}
\caption{ (a) The valley splitting as a function of electric field for monolayer MnBr. (b) The global band gap as a function of electric field. (c) The band structure of Mn layer resolved with SOC at E = 0.3 V/$\rm \AA$. (d) Schematic diagram of spin-layer locked anomalous valley Hall effect in the hole-doped monolayer MnBr at the K and K' valleys, respectively. The holes are represented by the + symbol. Downward arrows on behalf of the spin down carriers, respectively.
}
\end{center}
\end{figure}

\subsection{Built-in electric field tune valley splitting}
The successful preparation of Janus structure enriches the intrinsic physical properties of 2D materials \cite{43}. Since the upper and lower surfaces of the Janus structure are not equivalent, a large built-in electric field is generated. Interestingly, our previous investigation found that the built-in electric field can effectively tune the band structure and topological phase transition \cite{15}. By replacing the upper atomic layer of Br with Cl or I, as shown in Fig. 9(a,c) and Fig. S4(a, c), the space group and the point group are reduced from P$\bar{3}$m1 and D$_{3d}$ to P3m1 and C$_{3v}$, respectively. Our calculation results indicate that the energy of AFM1 state is 4.138 eV (4.804 eV), 3.205 eV (3.585 eV), and 2.146 eV (2.507 eV) lower than the FM, AFM2, and AFM3 states for Mn$_2$ClBr and Mn$_2$BrI, respectively.

\begin{figure}[htb]
\begin{center}
\includegraphics[angle=0,width=1.0\linewidth]{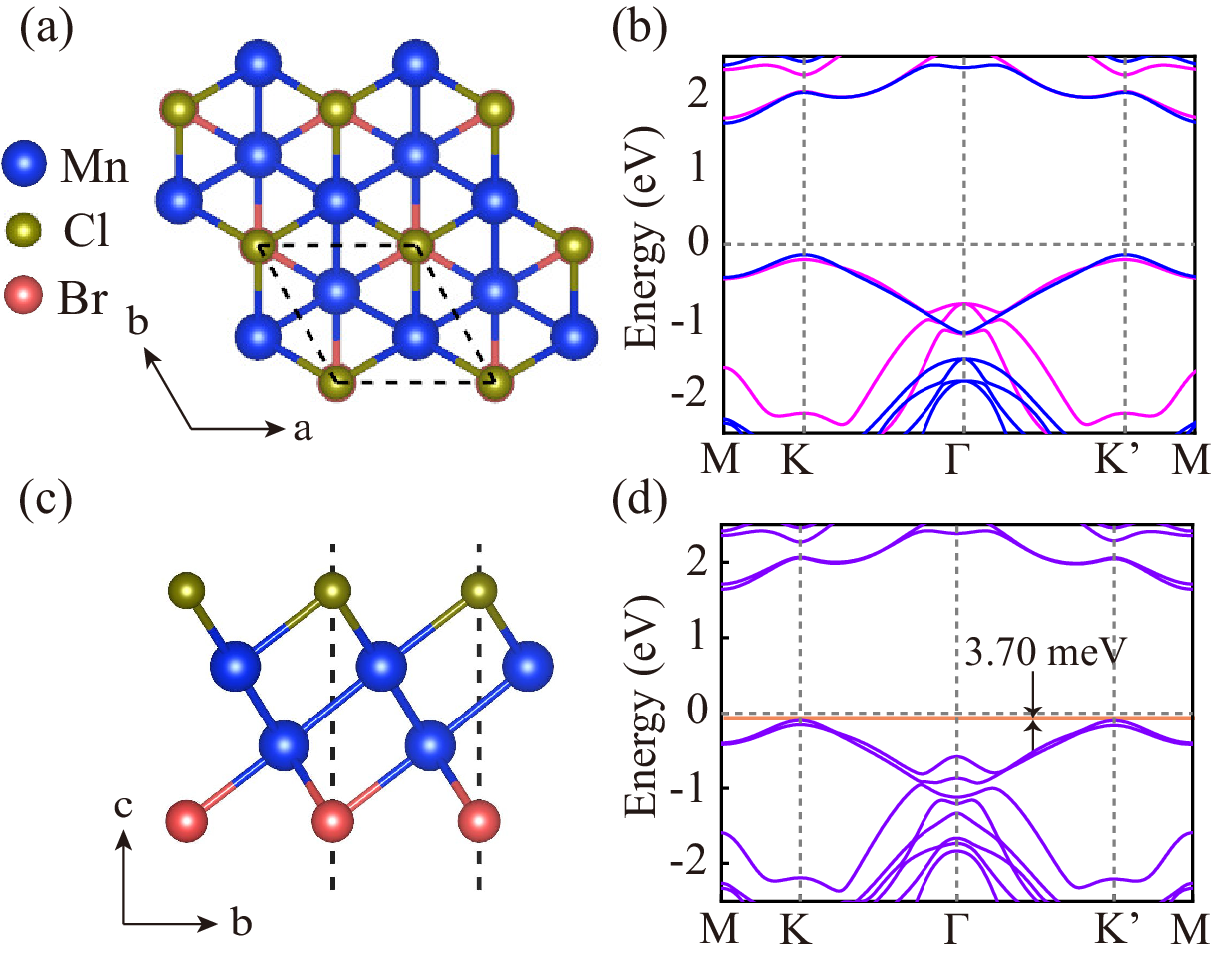}
\caption{ (a, c) The top and side views of the crystal structure for monolayer Mn$_2$ClBr, respectively. The blue, brown, and pink balls represent Mn, Cl, and Br elements, respectively. (b) Spin-polarized band structure of monolayer Mn$_2$ClBr. The blue and magenta lines denote spin up and spin down bands, respectively. (d) Band structure with considering the SOC. The valence band valley splitting is shown by the orange shading.
}
\end{center}
\end{figure}

In order to quantitatively analyze the effect of built-in electric field on valley splitting, we define the built-in electric field as E$_{in}$ = ($\Phi_2$ - $\Phi_1$)/$\Delta h$. The $\Phi_1$ and $\Phi_2$ denote the electrostatic potential at the top and bottom of Mn$_2$ClBr and Mn$_2$BrI, respectively. The $\Delta h$ is the structural height of Mn$_2$ClBr and Mn$_2$BrI. According to the plane averaged electrostatic potential of monolayer Mn$_2$ClBr and Mn$_2$BrI along the z axis, the electrical polarization can be clearly observed. Therefore, the built-in electric field is 0.04 V/$\rm \AA$ (Mn$_2$ClBr), and 0.07 V/$\rm \AA$ (Mn$_2$BrI). In the absence of SOC, the built-in electric field produces the same effect as the electric field, resulting in spin splitting [see Fig. 9(b) and Fig. S4(b)]. Unlike the electric field, the built-in electric field raises the energy level of the valence $\Gamma$ point. The VBM of Mn$_2$ClBr remains the K/K' point, while the VBM of Mn$_2$BrI becomes the $\Gamma$ point. Importantly, the built-in electric field of monolayer Mn$_2$ClBr reduces the valley splitting to 3.70 meV [see Fig. 9(d)]. It's worth noting that the spin splitting can also be achieved with a built-in electric field. Therefore, the spin-layer locked anomalous valley Hall effect can also be realized. Hence, it is further confirmed that the built-in electric field caused by the quantum confined effect can effectively tune the physical properties.

\subsection{Ferroelectric substrate tune valley splitting}
\begin{figure}[htb]
\begin{center}
\includegraphics[angle=0,width=1.0\linewidth]{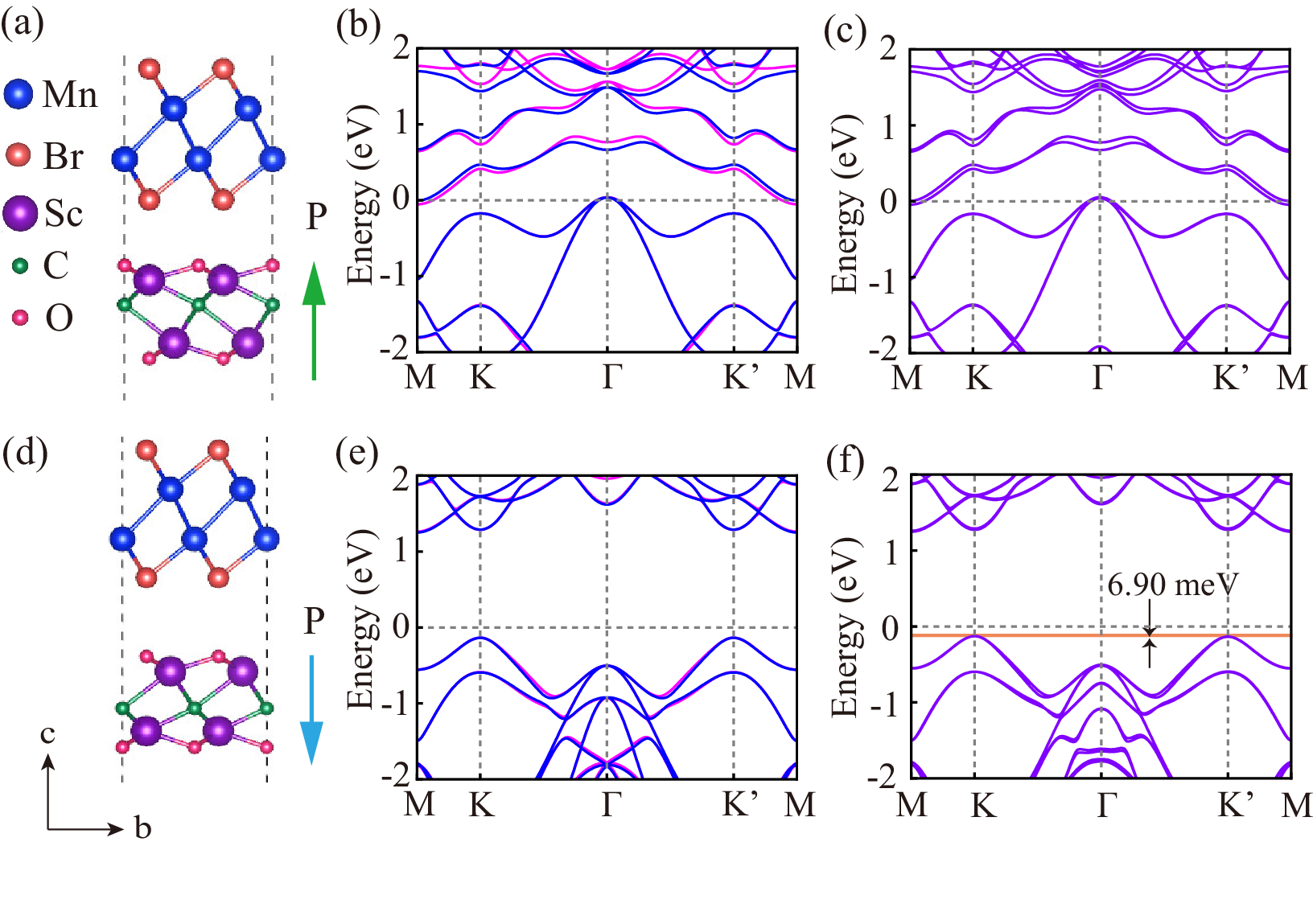}
\caption{ Side views of (a) P$\uparrow$ state and (d) P$\downarrow$ state MnBr/Sc$_2$CO$_2$ heterostructures with the hollow configuration. (b, e) Spin-polarized band structure of MnBr/Sc$_2$CO$_2$ heterostructures for (b) polarization up and (e) polarization down with the hollow configuration, respectively. The blue and magenta lines denote spin up and spin down bands, respectively. (c, f) Band structure with the SOC of MnBr/Sc$_2$CO$_2$ heterostructures for (c) polarization up and (f) polarization down with the hollow configuration, respectively.
}
\end{center}
\end{figure}

Ferroelectric substrate is also one of the common means of tuning physical quantities. Therefore, we explore the ferroelectric substrate Sc$_2$CO$_2$ tune valley splitting of monolayer MnBr. We found that the 1 $\times$ 1 unit cell of MnBr is matched to the 1 $\times$ 1 Sc$_2$CO$_2$ with a lattice mismatch rate of less than 5 $\%$. Here, we consider three typical configurations, as shown in Fig. 10(a, d), Fig. S6(a, d), and Fig. S7(a, d), i.e., the lower layer Mn atom of MnBr being in hollow, top-Sc, and top-O positions of the Sc$_2$CO$_2$, respectively. As listed in Table SI, the total energy results indicate that the hollow and top-Sc configurations are the most stable for polarization down and polarization up, respectively. Besides, the easy magnetization direction of all structures is along out-of-plane, as listed in Table SII. As the Sc$_2$CO$_2$ is polarized P$\uparrow$, as shown in Fig. 10(b), Fig. S6(b), and Fig. S7(b), the spin-polarized band structures show metallic properties. On the contrary, the band gap of heterostructure is $\sim$ 1.40 eV for the polarization P$\downarrow$ of Sc$_2$CO$_2$ [see Fig. 10(e), Fig. S6(e), and Fig. S7(e)]. Importantly, the VBM remains at K/K' point, and the K and K' in energy degenerate. When the SOC is included, band structures of P$\uparrow$ heterostructure show metallic properties, as shown in Fig. 10(c), Fig. S6(c), and Fig. S7(c). Interestingly, the valley degeneracy of K and K' disappears, and they exhibit $\sim$ 7.00 meV valley splitting [see Fig. 10(f), Fig. S6(f), and Fig. S7(f)]. It indicates that the transition from metal phase to valley polarization semiconductor can be realized when the direction of ferroelectric polarization switch.

\section{CONCLUSION}
In summary, based on the DFT and effective Hamiltonian model, we predict that antiferromagnetic monolayer MnBr is a ferrovalley material and realize the anomalous valley Hall effect. The valley splitting of the VBM is 21.55 meV at K and K' points. Importantly, the valley splitting can be effectively regulated by onsite correlation, strain, magnetization rotation, electric field, and built-in electric field. Importantly, the electric field and built-in electric field induce spin splitting due to breaking the P symmetry. Therefore, the spin-layer locked anomalous valley Hall effect can be observed in MnBr. More interestingly, when the ferroelectric substrate Sc$_2$CO$_2$ is the P$\uparrow$ state, the heterostructure shows a metallic property. However, while the electric polarization of Sc$_2$CO$_2$ switches to P$\downarrow$ state, the heterostructure becomes the valley polarization semiconductor. Our work not only enriches the valley physics, but also provides a variety of ways to tune valley splitting.

\section*{ACKNOWLEDGEMENTS}
This work is supported by the National Natural Science Foundation of China (Grants No. 12474238, and No. 12004295), the Natural Science Foundation of Guizhou Provincial Education Department of China (Grant No. ZK[2022]558), P. Li also acknowledge supports from the China's Postdoctoral Science Foundation funded project (Grant No. 2022M722547), the Fundamental Research Funds for the Central Universities (xxj03202205), the Open Project of State Key Laboratory of Surface Physics (No. KF2024$\_$02), and the Open Project of State Key Laboratory of Silicon and Advanced Semiconductor Materials (No. SKL2024-10). Y. She acknowledge supports from the NSF of Tongren Science and Technology Bureau (Grants No. [2023]41).


\end{document}